\documentclass[aps,prl,twocolumn,superscriptaddress]{revtex4}
\def\be{\begin{equation}}
\def\ee{\end{equation}}
\def\bea{\begin{eqnarray}}          
\def\eea{\end{eqnarray}}
\def\bi{\begin{itemize}}
\def\ei{\end{itemize}}

\usepackage{graphicx}
\usepackage{color}

\begin{document}

\title{ Comment on ``Quantum entangled dark solitons formed by ultracold atoms in optical lattices'' }

\author{Jacek Dziarmaga} 
\affiliation{Instytut Fizyki im. Mariana Smoluchowskiego and
Mark Kac Complex Systems Research Center, 
Uniwersytet Jagiello\'nski, ul. Reymonta 4, PL-30-059 Krak\'ow, Poland}

\author{Piotr Deuar} 
\affiliation{Institute of Physics, Polish Academy of Sciences, Al. Lotnik\'ow
32/46, 02-668 Warsaw, Poland}

\author{Krzysztof Sacha} 
\affiliation{Instytut Fizyki im. Mariana Smoluchowskiego and
Mark Kac Complex Systems Research Center, 
Uniwersytet Jagiello\'nski, ul. Reymonta 4, PL-30-059 Krak\'ow, Poland}

\maketitle


The recent Letter \cite{Carr} 
describes full quantum simulations of 
a dark soliton in a Bose-Einstein condensate in a regime where the system
cannot be described by the perturbative approach. The authors argue, 
based on the filling in of the two-point correlator $g^{(2)}$, 
that a photograph of a condensate would reveal a smooth atomic 
density without any localized dark soliton.
This is in contrast to the perturbative regime where a photograph would show a dark soliton with 
a random position \cite{DzS}. 
While we admire the quantum simulations and other results in \cite{Carr}, 
we think that their conclusion about the outcome of a single experiment is not justified 
by this property of $g^{(2)}$. 

As the question of whether and when dark solitons fill in in single realisations has been a 
recurring and sometimes confusing one in the field, it is important to pin down what conclusions can or cannot be made.
With this aim, we provide the following counterexample in the non-perturbative regime 
(see also \cite{nonpert}).

Let $\phi_q(x)\propto\tanh[(x-q)/\xi]$ be a standard condensate wave function with a dark soliton at $q$. 
With $\hat a_q=\int dx~\phi_q^*(x)\hat\Psi(x)$, a state
$\left(\hat a^\dag_q\right)^N|0\rangle$ is a condensate with a soliton at $q$, and let 
the $N$-particle state be a superposition of condensates with different $q$ 
\be
|\psi_0\rangle\propto
\int dq~
\psi_0(q)~
\left(\hat a^\dag_q\right)^N
|0\rangle~,
\label{psi0}
\ee 
where $\psi_0(q)$ defines the superposition.  

After measurements of $n$ atomic positions $x_1,...,x_{n}$ the state (\ref{psi0}) collapses 
to a conditional state
$$
|\psi_n\rangle ~\propto~
\hat\Psi(x_n)..\hat\Psi(x_1)
|\psi_0\rangle ~\propto~
\int dq~
\psi_n(q)~
\left(\hat a^\dag_q\right)^{N-n}
|0\rangle~,
\label{psin}
$$
where 
$
\psi_n(q)=\phi_q(x_n)..\phi_q(x_1)\psi_0(q)
$. The $(n+1)$-st measurement will find a particle
at $x_{n+1}$ with a probability 
$$
p_{n+1}(x_{n+1}) \propto \langle\psi_n|\hat\Psi^{\dagger}(x_{n+1})\hat\Psi(x_{n+1})|\psi_n\rangle. 
$$
This is equivalent to simultaneous measurement of all $x_i$.

Using the methods of \cite{DzS}, we simulated measurement of all $N=5000$ particles on 
a lattice of 31 sites, 
where $x,q\in\{-15,15\}$, assuming a soliton width $\xi=1.5$, and a delocalized 
(uniform) superposition $\psi_0(q)\propto1$ that is non-perturbatively wider than 
the soliton width. The inset in the figure shows the ensemble average particle 
density $p_1(x)$ and the main figure shows a generic histogram of particle positions 
$x_1,..,x_N$ measured in a single realization. Each single realization of 
the experiment finds a soliton localized at some definite but random $q$.

\begin{figure}
\centering
\includegraphics*[width=0.9\linewidth,clip]{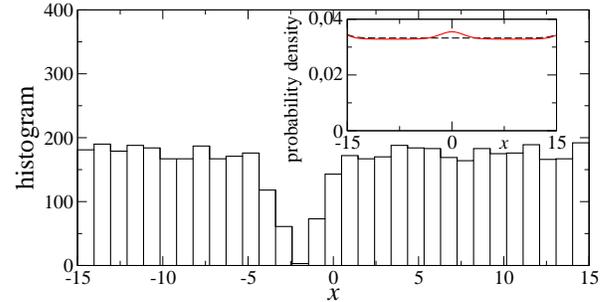}
\vspace*{-0.3cm}
\caption{(Color online) Histogram of measured atom positions in a single experiment from (\ref{psi0}). Inset: single
particle density $p_1(x)$ (dashed black line) and $g_2(x)$ (solid red line), normalized so that $\int g_2(x) dx = 1$.\label{one}}
\end{figure}

What about the two-point correlator
$g_2(x)=\langle\psi_0|\hat\Psi^\dag (0)\hat\Psi^\dag (x)\hat\Psi(x)\hat\Psi(0)|\psi_0\rangle$ that 
is analyzed in \cite{Carr}? This turns out nearly uniform (see the inset). Hence, the ``filled in $g_2$'' $\to$ ``filled in soliton'' 
line of reasoning is clearly incorrect. Of course, the soliton simulated in \cite{Carr} may still be greying, but the point here is that one cannot answer such a question by analyzing $g_2(x)$.

Our example demonstrates that, in some cases, a low order 
correlator like $g_2(x)$ is insufficient to draw conclusions on the outcome of a single experiment, and that the soliton in a BEC is such a case. 
Here, $g_2$ is equal to our $p_2(x_2)$ after the first particle was measured 
at $x_1=0$. A measurement of only one particle is not enough to collapse the soliton 
position. If we want to infer the soliton position from a histogram of particle positions, 
then the number of measured particles must be large enough to provide a histogram 
with a well-resolved soliton notch. This is a fundamental requirement for an accurate 
Bayesian inference of the soliton position.

Support within Polish Government scientific funds (2008-2011 KS, 2009-2012 JD) as a research 
project is acknowledged.


\begin{thebibliography}{9}

\bibitem{Carr} R. V. Mishmash and L. D. Carr, 
                  Phys. Rev. Lett. {\bf 103}, 140403 (2009); see also:
                  R. V. Mishmash {\it et al.}, 
                  Phys. Rev. A {\bf 80}, 053612 (2009).

\bibitem{DzS} J. Dziarmaga and K. Sacha, J. Phys. B {\bf 39}, 57 (2006);
              Phys. Rev. A {\bf 66}, 043620 (2002); 
              J. Dziarmaga {\it et al.}, 
              J. Phys. B {\bf 36}, 1217 (2003); 
              Phys. Rev. A {\bf 66}, 043615 (2002).
\bibitem{nonpert} J. Dziarmaga, Phys. Rev. A {\bf 70}, 063616 (2004).

\end{thebibliography}
\end{document}